# Malleable patterns from the evaporation of colloidal liquid bridge: *coffee ring* to *the scallop shell*

Ankur Chattopadhyay[1], Srinivas Rao S [1,†], Omkar Hegde[1,†], Saptarshi Basu[1,*]

[1] Department of Mechanical Engineering, Indian Institute of Science

*Corresponding author email: sbasu@iisc.ac.in

[†]These authors contributed equally to the work

## Abstract

The present article highlights an approach to generate contrasting patterns from drying droplets in a liquid bridge configuration, different from well-known coffee rings. Reduction of the confinement distance (the gap between the solid surfaces) leads to systematized nano-particle agglomeration yielding to spokes-like patterns similar to those found on scallop shells instead of circumferential edge deposition. Alteration of the confinement length modulates the curvature that entails variations in the evaporation flux across the liquid-vapor interface. Consequently, flow inside different liquid bridges (LBs) varies significantly for different confinement lengths. Small confinement lengths result in the stick-slip motion of squeezed liquid bridges. On the contrary, the stretched LBs exhibit pinned contact lines. We decipher a proposition that a drying liquid thin film (height ~ $O(10^{-7})$m) present during dewetting near the three-phase contact line is responsible for the aligned deposition of particles. The confinement distance determines the height of this thin film, and its theoretical estimations are validated against the experimental observations using reflection interferometry, further exhibiting good agreement (in order of magnitude). Modulating the particle size (by ~100's of nanometers) does not significantly influence the precipitate patterns; however, particle concentration can substantially affect the deposition patterns. The differences in deposition patterns are attributed to the complex interplay of the gradient of evaporation flux induced motion of contact line in combination with the drying of thin liquid film during dewetting.

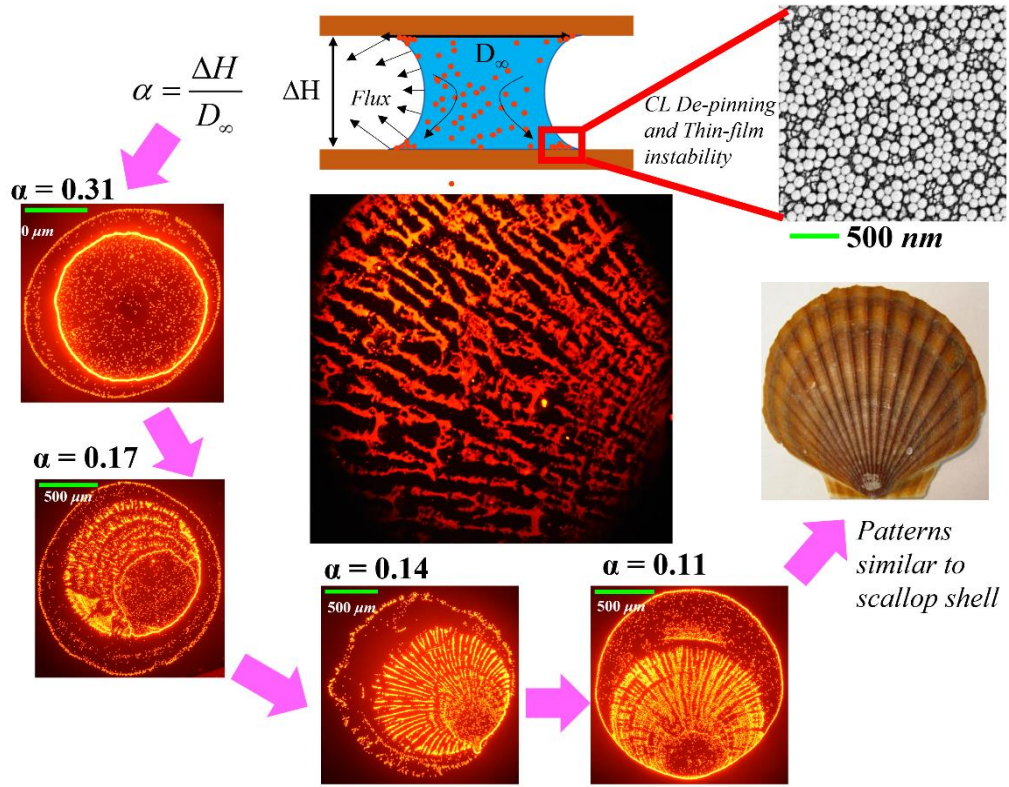

**Abstract Figure**

# 1. Introduction:

The formation of patterned drying precipitates from evaporating colloidal drops is a complex non-equilibrium process, where the multiscale heat and mass transfer of the droplet are governed by the coupled interplay between solvent evaporation and advection−diffusion of particles. Several factors, such as surrounding atmosphere (temperature, humidity), substrate properties (wettability, substrate temperature), and nature of particle (for example, particle size and shape), can be tailored to regulate the outcome of the deposition patterns that finds potential applications in ink-jet printing [1], fabrication of functional coatings [2–4] and development of miniaturized electronic devices [5]. The seminal article by Deegan et al. [6] reported that evaporation of a sessile droplet containing particles typically leads to a ring-like deposit, commonly referred to as ''coffee-ring''. In this case, the droplet contact line (CL) remains pinned to the substrate, and non-uniform evaporative flux along the liquid-vapor interface generates radially outward capillary flow that advects the particles towards the pinned edge, thereby resulting such deposits. Variations in deposit patterns, for instance, multiple rings instead of a single one, central deposits, hexagonal cells, and even uniform deposits, were also observed by fellow researchers [7-9]. Further, depinning of the CL and Marangoni stresses has been found to determine the precipitation patterns. The contribution of thermal Marangoni advection induced by the latent heat of evaporation [10] or activated due to substrate heating [11] on the deposition dynamics is well documented. Recent reports have shown that inducing differential vapor gradients around droplet surroundings leads to a reconfiguration of CL that modifies the residual precipitates [12, 13]. Therefore, simple yet intriguing patterns can be obtained by manipulating CL, controlling the evaporative flux across the droplet interface, and modulating the associated fluid flow within the evaporating drop.

Researchers from different backgrounds have investigated the evaporation dynamics of a sessile drop and the subsequent emergent patterns due to solvent depletion. However, very little attention is paid to drying droplets in the capillary liquid bridge configuration, particularly in the case of colloidal dispersions. In many biological and industrial processes [14, 15], this type of confined drying of droplets is observed that offers an advantage in controlling the solvent evaporation flux so that engineered deposition patterns can be realized. One of the earliest reports by Lin and Granick [16] on colloidal deposits for restricted geometry (sphere-on-plate) demonstrated the formation of concentric rings. Similar observations (concentric rings with gradients) were also confirmed by Xu et al. [17] when they explored the drying of liquid bridges under sphere-on-plate geometry. Considering sphere-on-plate configuration, Mondal and Basavaraj [18] concluded that the concentration of the particles influences deposition patterns of suspensions containing ellipsoidal particles. The sphere-on-plate geometry is not exactly symmetric as one end (top) of the liquid bridge is attached to a curved surface, whereas the other end (bottom) with a flat surface. This configuration inherently induces contact angle (CA) variations at both ends. This may contribute to the differences in evaporation flux across the meniscus of the droplet due to the geometrical restrictions. These effects can be mitigated if one chooses a parallel plate configuration.

Only a handful of research groups have attempted to study the evaporation of colloidal drops confined by parallel plates. Experiments performed with concentrated colloids (very high volume fractions) exhibit distinct deposit patterns formed due to interfacial instability [19] or buckling phenomenon [20]. Mahanta and Khandekar [21] compared the evaporation characteristics of liquid bridges (pure solvent and nano-particle (CuO) based media) between the heated surfaces and concluded that the presence of particles enhances the overall rate of evaporation owing to higher diffusional conductivity. Of late, the article by Mondal and Basavaraj [22] demonstrated the role of the confinement length (distance between the solid surfaces) and particle concentration on the deposits. According to them, the continuous stick-slip motion of the CL resulted in symmetric spiral deposits independent of the shape of the particles. Very recently, Upadhyay and Bhardwaj [23] examined the colloidal liquid bridge by altering the wettability of solid surfaces and the size of particles along with the composition of the colloids. Their observations include the formation of multiple rings governed by the stick−slip motion of CL. Besides, they reported notable asymmetric deposits composed of larger particles, with the majority of the particles present at the bottom surfaces as their transport is influenced by gravitational forces rather than the hydrodynamic forces.

From the discussion of the literature survey, it is understood that evaporation of colloidal liquid bridge can result in varying deposition patterns that differ from the conventional "coffee ring". This opens up an opportunity to tailor the precipitate patterns that may be of greater interest to several industrial and biological processes. Moreover, one must comprehensively understand the fundamental insights into the CL dynamics during evaporation and the physical deposition mechanisms of particles that yield a specific pattern. Following this aim, we study the evaporation of colloidal liquid bridges (LB) in parallel plate configuration. In this contribution, we experimentally demonstrate how the commonly observed "coffee ring" can be turned into differently orientated patterns (similar to spokes) by modifying only the confinement length for the same particle concentration. To the best of our knowledge, no article has reported this kind of pattern in the case of colloidal liquid bridges. The drying kinetics of the colloidal LB systems and the role of specific parameters in controlling those patterns are discussed thoroughly. It has been shown that the evaporation flux across the surface of LB is strongly dependent on the length of the confinement and the curvature of the surface of LB. The evaporation flux across the surface of the LB is a deciding parameter in the depinning of CL for lower confinement lengths leading to thin-film instabilities and unique pattern formation. The thin-film theory explains the mechanism that led to the differences in deposition patterns, and further interferometric measurements were carried out that confirm the theoretical proposition.

## 2. Materials and Methods:

### 2A. Materials

The colloids were prepared by suspending neutrally buoyant, monodispersed, charge stabilized, fluorescent polystyrene particles (purchased from Thermo Fisher Scientific) to de-ionized (DI) water. The suspensions were sonicated for about 30 mins, resulting in a homogeneous distribution of particles in the liquid. Post sonication, the suspensions were stored in centrifuge tubes (Tarsons Product Ltd.). The particles do not form sediment or agglomerate for at least two weeks from preparation (refer to the zeta potential data of the particles in the supplementary information for stability). Three different colloids with varying concentrations ($\varphi$ = 0.02%, 0.1%, 1% v/v) were prepared for monodispersed particle size of 860 ± 20 nm and 100 ± 20 nm having 1% volume fraction. The substrate used in the present study were the glass slides of dimensions 25 × 25 × 1 (length × breadth × height) mm$^3$. The standard glass slides with a planar area of 75 × 25 mm$^2$ (procured from Blue Starc) were meticulously sliced with a glass cutter to obtain the required dimension. The square slides were designed based on the idea of minimizing the edge effects so that a drop placed at the center of that slide is almost equidistant from both the spatial dimensions. These square slides were sonicated in a bath of propan-2-ol (isopropanol) for 10 mins and further rinsed by Kimwipes (Kimberly Clark International). All the experiments were performed by maintaining the temperature of 23–25 °C and relative humidity around 45% ± 2% (measured with a sensor TSP-01, provided by Thorlabs).

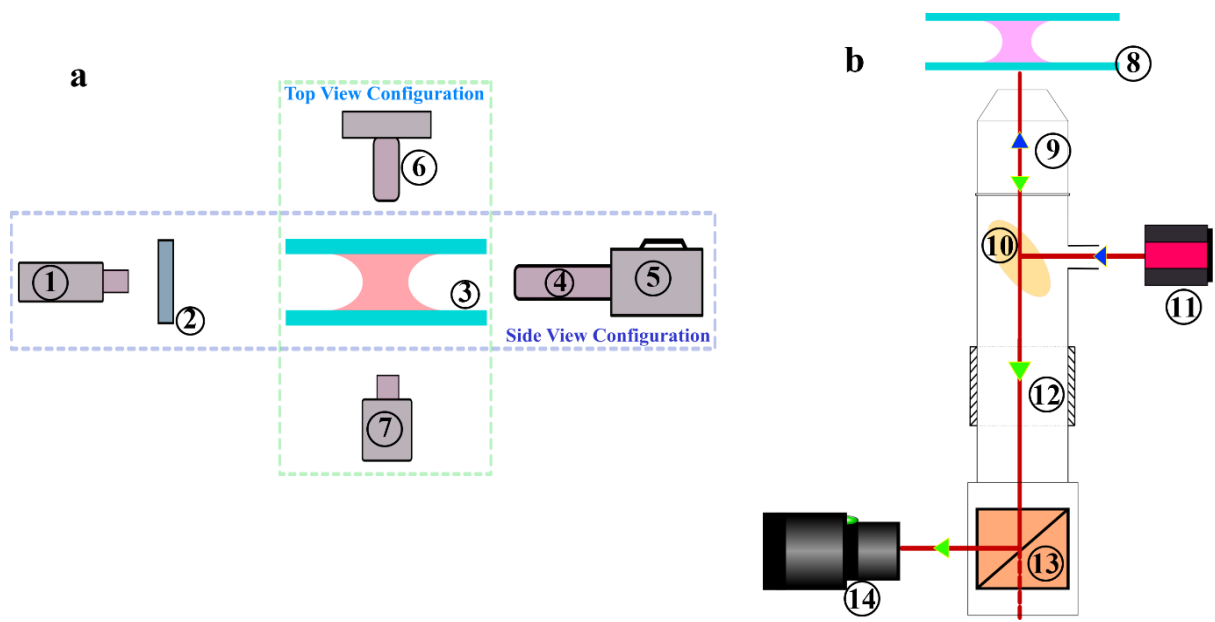

**Figure 1.** Schematic of the experimental set-up – (a) Various components used in Shadowgraph and Fluorescent Microscopy: (1) Light source, (2) Diffuser, (3) Liquid Bridge, (4) Focus/zoom lens assembly, (5) Digital camera, (6) Camera attached to the microscope, (7) Integrated light source of the microscope. (b) Different items to perform reflection interference

microscopy: (8) Liquid Bridge, (9) Microscopic objective, (10) Beam splitter (11) Pulse-laser source (640nm), (12) lens assembly, (13) Di-choric mirror, (14) High-speed camera.

2B. Experimental set-up

The colloidal droplet of $0.3 \pm 0.1$ μl was minutely placed on to the central part of the square slide (bottom) using a micropipette (purchased from Thermo Scientific Finnpipette, range: 0.2–2 μl) and later the top slide was positioned accordingly with the help of an actuator, so that drop was allowed to dry in liquid bridge (LB) configuration. The evaporation dynamics of the sandwiched droplet were recorded at 30 fps from the side view using a digital camera (Nikon D5600) fitted with a zoom lens assembly (Navitar) in the presence of a controllable LED light source (5W, procured from Holmarc), as shown in fig. 1a. A diffuser was placed between the light source and the LB for homogeneous light distribution. An optical microscope (Olympus) fitted with a CCD camera (Nikon D7200) simultaneously visualized the contact line (CL) dynamics of the LB from the top. The dried precipitates were volumetrically illuminated using a mercury source with a filter, and the emission signal (612 nm) was captured (at $10\times$ to $50\times$ magnification) by the same microscopic arrangement described earlier. Image analysis was carried out with open source software FIJI (ImageJ). Besides, the deposition patterns were also scrutinized by SEM to observe the particle aggregation and analyzed with the help of an optical profilometer (Talysurf Hobson) to estimate the thickness of the dried colloids.

Evaporation dynamics (dewetting phenomena of the liquid layer) of the colloidal LB at variable confinement heights and their temporal deposition patterns were examined with the help of reflection interference microscopy. Microscopic alignment and the arrangement of associated components are shown in fig.1b. A pulse-laser source of 640 nm wavelength (*Cavitar, Finland*) of the light source is positioned on a beam splitter that transmits and reflects the light beam. The reflected light beam passes through the microscopic objective (*4X* and *20X*) and reflects from the bottom glass plate positioned at the focal length of the microscope. The developed phase difference of the light beam at the leading of the LB repels from the glass surface and passes through the objective, beam splitter, zoom lens (Navitar), di-choric mirror, as seen in fig. 1b. The change in phase difference induces the interference fringe patterns that are captured using a high-speed camera (Photron SA-5, LaVision). The detailed procedure of postprocessing the images obtained from these experimental observations is included in the supplementary section.

## 3. Results and discussions:

3A. Formation of the liquid bridge and its lifetime

A sessile aqueous drop of dispersed particles after evaporation of solvent leave deposits of particles. In most of these dried precipitates, the particles accumulate at the three-phase contact line (CL), forming well-known ''coffee-ring'' patterns. However, contrasting patterns emerge when a colloidal drop is allowed to evaporate in a liquid bridge (LB) configuration, confined between two impermeable solid surfaces. In this context, before proceeding to the analysis of deposit patterns, we will discuss some of the important chronological events (fig. 2a) of a liquid bridge (LB), starting from its inception to the final evaporation. Initially, a colloidal sessile drop was gently placed on a glass slide, and then, with the help of a position-controlled actuator, we brought down another glass slide on top of the droplet (refer to the fig. 2b). When the top surface touched the sessile drop, it rearranged itself into the 'liquid bridge' (LB) configuration. Next, the 'confinement length' ($\Delta H$) (the distance between two parallel surfaces) was adjusted to a suitable height, and the same confinement length was maintained throughout the evaporation process. The time taken by a sessile drop to attain stable LB configuration is designated by $t_{ib}$ (fig. 2a). As the evaporation continued, the drop lost more and more solvent as it remained confined between two parallel solid surfaces. The neck in the central region of the LB became narrow with evaporation, and later, the LB was divided into two individual drops. Further, those two drops underwent evaporation until final precipitates formed on the top and bottom glass slides. The period during which the drop remains in its LB configuration while undergoing evaporation before its separation into individual drops is referred to as $t_{lb}$. The evaporation timescale of the individual droplets from their evolution to completion of drying is given a symbol $t_{pb}$. The addition of these three-time scales yields the lifetime of the drop. The contact angle (CA) of the sessile colloidal drop was observed to be 33°±5°, and the respective initial contact diameter ($D_{s0}$) was measured to be ~ 1.8 mm. A variable '$\alpha$' denotes the non-dimensional confinement distance, as $\Delta H$ is normalized with respect to $D_{s0}$. The higher the $\alpha$, the larger is the gap between the two surfaces.

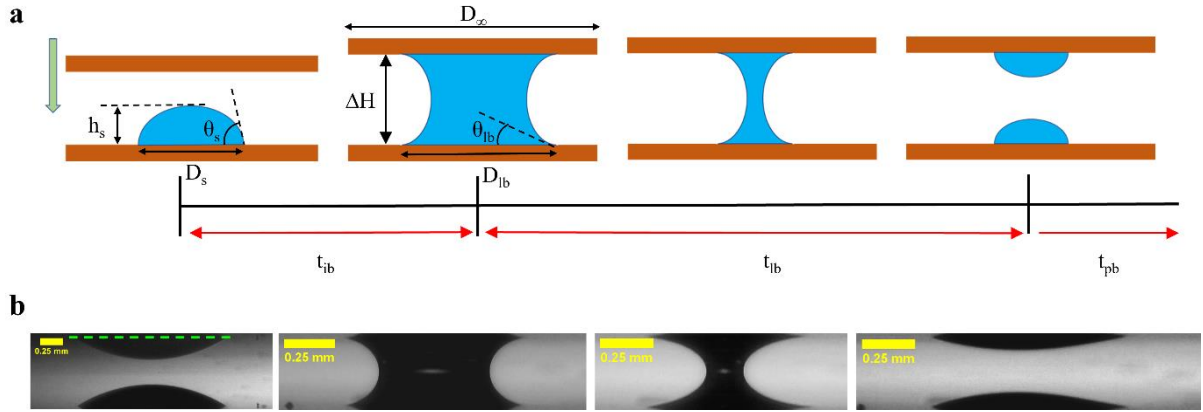

**Figure 2.** (a) Schematics chronology of the capillary liquid bridge. The time interval between placing a sessile drop on the bottom glass slide and achieving a stable LB configuration is designated as $t_{ib}$. The timespan till which the drop remains in its LB configuration while undergoing evaporation before separating into two individual drops is referred to as $t_{lb}$. The time when individual droplets undergo evaporation post separation until the final precipitates are formed is indicated by $t_{pb}$. When placed on a glass slide, the sessile drop contact diameter (CD) and droplet height are designated by $D_s$ and $h_s$, respectively. The distance between two parallel slides, referred to as 'confinement length', is indicated by $\Delta H$. The CD and CA in LB configuration are indicated by $D_{lb}$ and $\theta_{lb}$. The end-to-end distance of the square glass slide is represented by $D_\infty$. (b) The experimental images of confinement distance $\alpha = 0.2$ at corresponding time scales mentioned in fig. a. In the first image of fig. b, the dashed green line indicate the bottom edge of the top glass slide. The top droplet in the first image of fig. b is the fictitious reflection of the actual sessile droplet, placed at the bottom surface. However, the the final image in fig. b shows the individual drops after the separation of the LB.

### 3B. Patterns of colloidal liquid bridge

Figure 3a illustrates the final deposition patterns that emerged from colloidal liquid bridges. The confinement distances of those systems were altered ($\alpha = 0.11$ to $0.31$), keeping the same volume fraction ($\varphi = 1\%$ v/v) and particle size (860 nm). We further classify the confinement distance regimes into low ($\alpha < 0.2$), intermediate ($0.2 \leq \alpha \leq 0.3$) and high ($\alpha > 0.3$). The analysis of deposit patterns of bottom surfaces reveals an outermost "coffee ring" that is present for all the cases (refer to fig. 3a, enlarged view). This ring forms due to capillary flow when the sessile drop is placed initially on the glass surface (bottom one) as the droplet maintains its spherical cap configuration and remains pinned to the bottom surface. The droplet tries to attain the LB configuration when the sessile droplet comes in contact with the top surface. This transition from sessile drop to the liquid bridge is relatively faster, and the corresponding timescale is observed to be of the order of a few milliseconds. This action reduces contact diameter (CD) at the bottom surface as the drop attempts to conserve its volume. The shrinkage in CD at the bottom surface is complemented by the spreading of CD at the top surface. This is a highly non-equilibrium event and is accompanied by the transport of capillary waves as the droplet is in the process of transforming into its new LB configuration. While the sessile droplet attempts to attain stable LB configuration, the remaining thin liquid layer, present in the area between the initial CD (sessile) and new CD (LB) at the bottom surface, undergoes evaporation, leaving traces of colloidal dispersions (evident for all cases irrespective of modification in confinement distances, refer fig. 3a enlarged view – the region between the outer ring and central deposition pattern). The evaporation characteristics of this thin layer are addressed in detail in a later section (refer to section 3F). In this article, the deposition patterns refer to the corresponding central deposits unless otherwise specified.

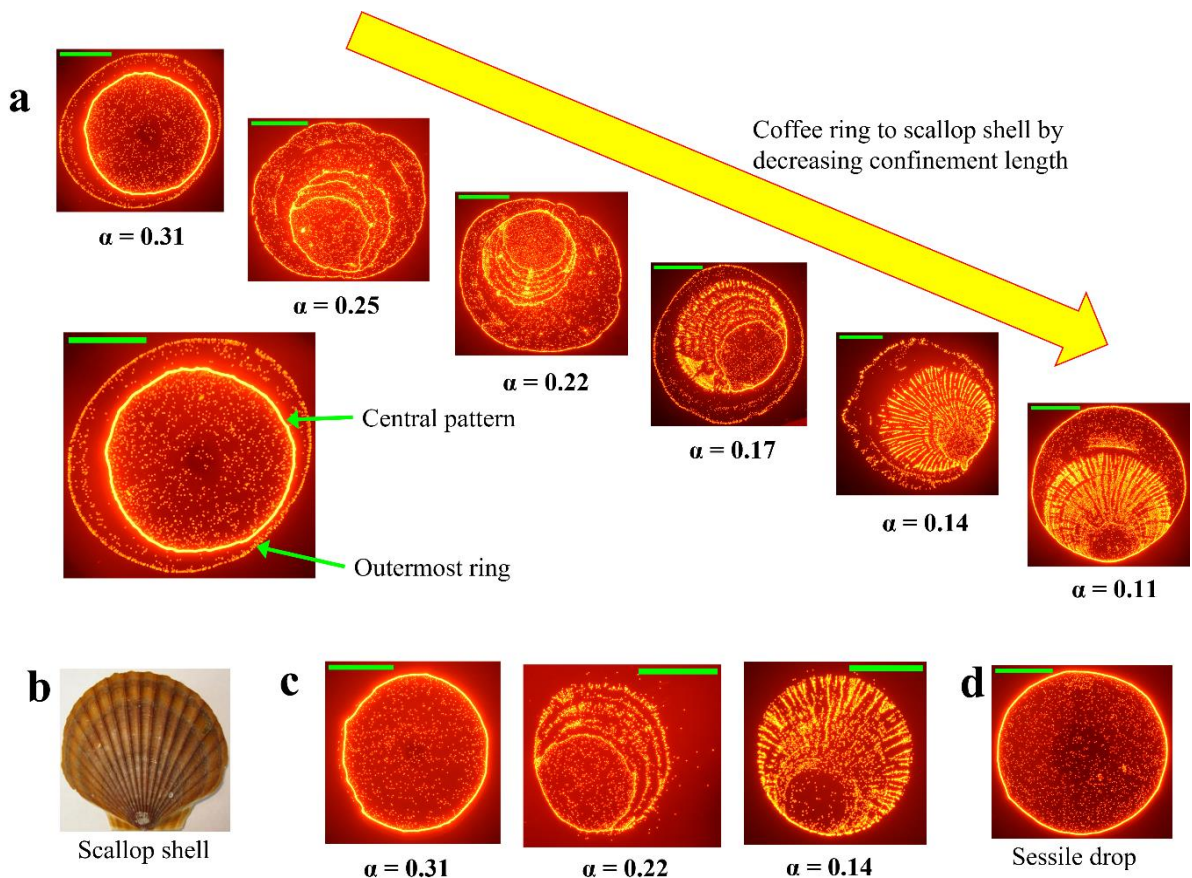

**Figure 3.** Varying deposit patterns of a colloidal (φ = 1% v/v, 860nm) liquid bridge. (a) Alteration of confinement distance (α varies from 0.11 to 0.31) leads to variation in precipitation patterns observed at bottom surfaces. At a higher confinement distance, coffee ring patterns are noticed. However, decreasing the gap leads to differences in patterns as particulates arrange themselves like spokes instead of the ring (for instance, α = 0.11). An enlarged view differentiates the outer ring from the central deposition pattern. (b) A typical figure of a scallop shell (copyright belongs to [24]). The central patterns at lower confinement distances (corresponding to α = 0.11, 0.14) have striking similarities with a scallop shell. (c) Deposit patterns at top surfaces of different confinement distances (α = 0.14, 0.22, 0.31). (d) Commonly observed "coffee ring" pattern. All the fluorescent images are acquired at 10× magnification. Scale bars, represented by the green line, are equal to 500 μm.

For higher α (α > 0.3), the LB remains pinned at both top and bottom surfaces. This behavior allows the particles to accumulate at the CL of LB, thereby resulting in a central coffee ring. In the case of intermediate and low confinement distances, the LB goes through a 'stick-slip' motion as it continuously depletes solvent (water) to the surroundings. However, there are considerable differences among the patterns of intermediate and lower cases (fig. 3a). While the intermediate LBs (α = 0.22, 0.25) exhibit multiple coffee rings, further reduction of confinement distance (α = 0.11, 0.14) leads to the evolution of 'spoke' like deposition patterns instead of coffee rings. For α = 0.17, spokes are observed; however, not continuous like the other two cases (α = 0.11, 0.14). These types are ordered particle alignment, analogous to spokes, are unprecedented and have not been reported till now to the best of authors' knowledge, particularly from the drying of colloidal LBs. We identify these spokes-like patterns as "scallop shell" patterns due to their peculiar similarity with the patterns observed on the outer shells of the scallops (fig. 3b). The corresponding precipitates at the top surfaces are replicas of the bottom surface deposits; however, the outermost ring is absent (fig. 3c). This is expected as the initial deposition of colloidal sessile drop on the bottom surface leads to the outermost ring due to capillary flow-induced particle aggregation at the initial CL, while the deposition at the top surface begins only after the LB is formed. The deposition patterns at the top surface also reveal similar sequences (from "coffee ring" to "scallop shell") as the confinement distance is diminished (α = 0.31 to 0.14).

To understand what led to the differences in patterns, we systematically investigate the evaporative behaviors of these systems. A sessile colloidal drop forms a single coffee ring by virtue of its loss of solvent around the three-dimensional surrounding space while the particles aggregate at its pinned CL. By sandwiching the droplet between two parallel solid surfaces, a part of its evaporation flux component is suppressed (that allows vertical mass transport from the sessile drop); while allowing its mass transfer to happen sideways, more so in the radial direction (refer to fig. 2b). To estimate the role of evaporation suppression in the pattern formation, we carried out additional experiments where we placed the top glass slide to the nearest possible vicinity of the apex of sessile drop (however, not in contact with the sessile drop), and the sessile drop remains unperturbed in the presence of the top surface. In this case, a very thin air gap existed between the droplet apex and top surface, and the height of this air layer was measured to be ~ $O(10^{-6})$ m. The initial droplet height ($h_{s0}$) was found to be sufficiently large ($O(10^{-4})$ m) than this air layer, higher by at least two orders of magnitude. The objective was to compare the performances of the sessile drop and LB by maintaining very similar magnitudes of confinement spaces ($\Delta H_{sessile}$ = 258 µm and $\Delta H_{LB}$ = 250 µm, in terms of α: $\alpha_{sessile}$ = 0.143, $\alpha_{LB}$ = 0.14) and to examine the role of confinement in their evaporation characteristics. After complete evaporation, the coffee ring pattern appears for the sessile case (fig. 3d), while the LB resulted in a scallop shell pattern (fig. 3c, $\alpha_{LB}$ = 0.14), as noted previously. This gives us an impression that the distribution of evaporation flux across the curvature of the drop, which governs the mass transport through the liquid-vapor interface, may play a domineering role in modifying its evaporation dynamics and in formation variant residual precipitates rather than the vertically imposed spatial constriction.

3C. Evaporation dynamics of colloidal liquid bridge

The salient features of evaporation of the colloidal LB are presented in fig. 4. As observed from fig. 4a, the temporal variation of the non-dimensional CDs ($D_{lb}^*$) at the bottom surface, obtained by normalizing the instantaneous CDs ($D_{lb}(t_{lb})$) of LB with respect to the corresponding initial CDs ($D_{lb0}$), is plotted. For low confinement distance (α = 0.11), the maximum reduction in CD is noticed. With an increasing gap, the decrement in CD is relatively low, and at the highest confinement distance (α = 0.34), hardly any change in CD is observed as the CL appears to be almost static. However, the evaporation behaviors of the squeezed LBs are entirely different from the higher α cases. In these cases, the mass transfer mechanism is governed by the stick-slip motion leading to a significant reduction in CD. One can notice a slight enhancement of CD (fig. 4a) during the penultimate stages (just before the rupture of LB, t ~ 0.95$t_{lb}$) of dewetting. This is attributed to the complex interplay among the dynamic change in evaporation flux across the curvature of the LB, the surface tension, and the gravitational forces that are responsible for splitting the LB into individual drops. The time-dependent variations of CDs corresponding to top surfaces show analogous behaviors to the bottom surfaces.

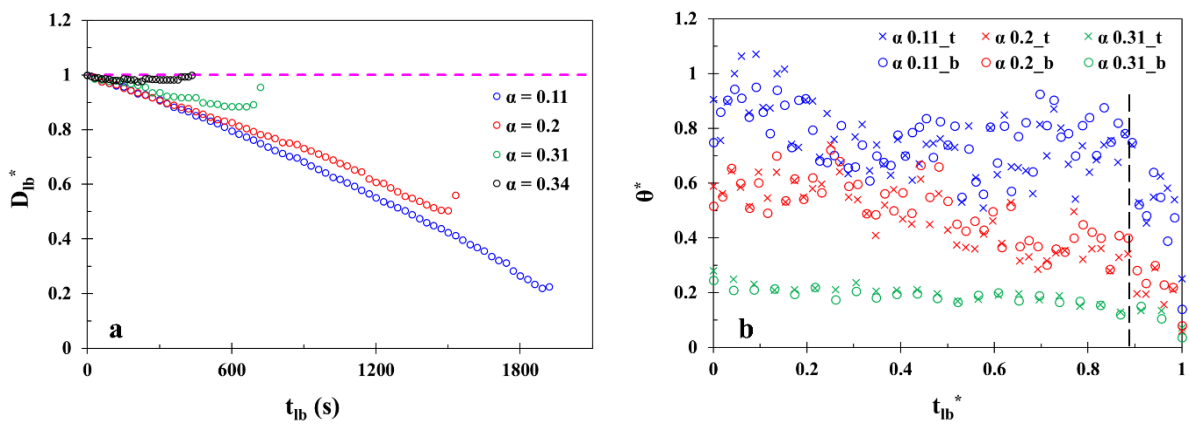

**Figure 4.** Evaporation kinetics of colloidal LB. (a) Temporal variation of non dimensional contact diameter ($D_{lb}^* = \dfrac{D_{lb}(t_{lb})}{D_{lb0}}$) with time ($t_{lb}$) for four different confinement lengths (α = 0.11, 0.2, 0.31, 0.34). $D_{lb}(t_{lb})$ represents the instantaneous contact diameter of the LB drop (refer fig. 2) and $D_{lb0}$ indicates the initial contact diameter of the drop after it attains stable LB configuration. (b) Dynamics of non dimensional contact angle (CA) ($\theta^* = \dfrac{\theta(t_{lb})}{\theta_{s0}}$) against non dimensional time ($t_{lb}^*$) (normalised against the individual $t_{lb}$) for three different confinement lengths (α = 0.11, 0.2, 0.31). $\theta(t_{lb})$ denotes the instantaneous CA of the LB drop, and $\theta_{s0}$ (33°) represents the mean initial CA of the colloidal drop in sessile configuration. The suffixes 't' and 'b' indicate the CAs corresponding to the top and bottom surfaces. The dashed pink line in fig. (a) indicates all the values corresponding to $D_{lb}^* = 1$. The dashed black line in fig. (b) represents a critical $t_{lb}^*$ that demarcates two zones. Gradual decrement in CA is observed in the left of this time while the slope changes are very sharp on its right side. This critical timescale

is estimated to be ~ 0.85-0.9 $t_{lb}^*$. The maximum mean error in estimating $D_{lb}^*$ is within a limit of ±3%, and the maximum uncertainty while analyzing CAs is ~ 2°.

Figure 4b compares the transient variations of CAs (top and bottom of the left curvature of corresponding LBs) of different confinement distances. The instantaneous CAs ($\theta(t_{lb})$) of LBs at top and bottom surfaces are normalized against the initial CA ($\theta_{s0}$) of the sessile drop to obtain the non-dimensional CAs ($\theta^*$). With time, the $\theta^*$ exhibits a declining trend upto a critical time threshold, beyond which the slope (the rate of change of CA with time) becomes steeper (by 6-10 times) (fig. 4b). This threshold (indicated by the black dashed line, fig. 4b) is noted to be ~ 0.85-0.9 $t_{lb}$, taking all the cases into account. As opposed to a pinned LB (α = 0.31) whose CA gradually decreases with time, in the case of the LBs showing considerable dewetting, the CAs show a non-monotonic trend. Between successive slipping movements, the LB adjusts to a new position, thereby contributing to temporal alteration of CAs. With increasing the gap between the surfaces, the magnitude of the initial CAs decreases. This is because, for a given volume of liquid, the LB with higher confinement space has to compensate for the droplet volume via the reduction in initial CD and CA compared to the more squeezed LBs. The differences between the CAs corresponding to the top and bottom surfaces for a given confinement length are estimated to be around a maximum of 7°. With increasing the confinement distance, this deviation becomes minimum.

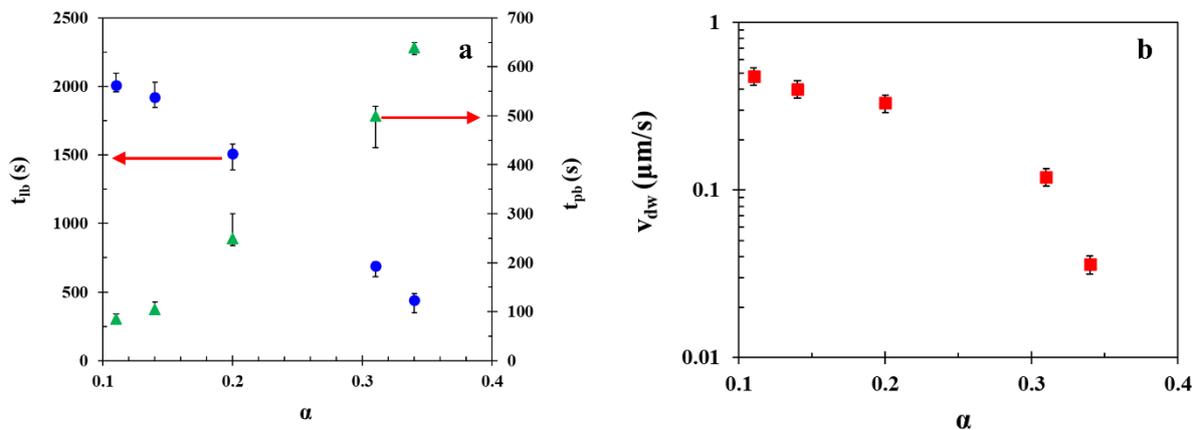

**Figure 5.** (a) Comparison of two timescales ($t_{lb}$ and $t_{pb}$) against different confinement length ($\alpha$ = 0.11 to 0.34). The blue circle corresponds to $t_{lb}$, and the green triangle represents $t_{pb}$. (b) Dewetting velocity as a function of confinement distance.

Further, we analyze the timescales and dewetting velocity as a function of the confinement distance (fig. 5). The detailed interpretation of these individual timespans is already explained in section 3A (refer to fig. 2). The $t_{ib}$ is estimated to be ~ 90 – 120 seconds depending on adjustments required to achieve desired confinement distance. With increasing confinement distance, the $t_{lb}$ decreases significantly. Analysis reveals three times enhancement in α results in ~ 75% reduction of $t_{lb}$ (fig. 5a), while $t_{lb}$ becomes larger. This is expected as more stretched (high α cases) LBs divide early and the droplets that emerge from these cases have larger volumes that require more time to evaporate than the lower confinement cases. The dewetting ($v_{dw}$) velocity is calculated by measuring the average displacement of CL over the entire timescale ($t_{lb}$) during which the drop remained in LB configuration. The estimated $v_{dw}$ demonstrates a non-monotonic decrement as it decreases by one order (from $O(10^{-7})$ m/s to $O(10^{-8})$ m/s) with increasing the confinement distance (fig. 5b). This is consistent with the characteristics portrayed in fig. 4a as a maximum reduction in CD was observed for the case having the highest dewetting velocity. A higher dewetting velocity can be related to the stick-slip motion of the LB drop, while low $v_{dw}$ indicates almost stationary CL.

Besides, we compare the experimentally observed $v_{dw}$ with the theoretical one (fig. 5b). To find out the same, we first evaluated the evaporation flux by using the following expression [23]

$$j_{th} = \frac{2D(1-RH)c_{sat}}{D_{lb0} \ln(D_\infty / D_{lb0})} \qquad (1)$$

In the eqn. 1, $D$, $RH$, $c_{sat}$, and $D_\infty$ represent mass diffusivity ($2.5 \times 10^{-5}$ m²/s) of liquid vapor in air, relative humidity (45%), saturated water vapor concentration (0.0232 kg/m³) at the vapor-liquid interface, the length scale far away from the squeezed drop (we consider the distance between two extreme ends of the glass slide, equal to 25 mm, refer fig. 2), respectively. The estimated $j_{th}$ comes about $1.35 \times 10^{-4}$ kg/m²s. The theoretical dewetting velocity (($v_{dw})_{th}$), calculated as per $(v_{dw})_{th} \sim j_{th}/\rho$ [25] is found to be ~ $O(10^{-7})$ m/s. This shows that theoretical dewetting velocity agrees with experimental values of lower confinement distances; however, it fails to accurately predict the velocity scales of LBs having higher confinements (α > 0.3). Since this expression is derived from the steady-state equation, this cannot capture the transient variation of evaporation flux. Moreover, this does not consider the curvature effects and role of confinement distances.. Although equation 1 has its limitations, the derived velocity scale from it still fairly estimate the experimental ones for low confinement distances.

3D. Evaporation induced flow inside the liquid bridge

Differential evaporation across the liquid-vapor interface of a sessile droplet induces a flow inside the droplet, commonly known as Evaporation Driven Flow (EDF) [6, 26]. Analogous to sessile droplets, evaporation occurs differentially over the surface due to the surface curvature of the LB [23]. The evaporation flux is maximum near the glass surface and minimum at the center of the LB as in the present study, for all cases, the meniscus is concave [23]. The LB surfaces have varying curvatures depending upon the value of α (also, the curvature changes dynamically with evaporation); evaporation flux is dependent on the curvature of the surface and can influence the flow inside the LB. It is thus crucial to understand the variation of the concentration field of water vapor across the surface of LB of different curvatures (i.e., different values of α) to understand its influence on flow inside the LB. Evaporation from the surface of the LB can be considered as quasi-steady, the diffusion equation ($\nabla^2 C = 0$) is the governing equation of the concentration field. The two-dimensional diffusion equation for the gas phase is solved using ANSYS FLUENT software (2021, R2). The curvature profile of the LB (for the fluid domain is as shown in fig. 6a) for different values of α is imported from the experimental data (images obtained from the shadowgraph). Following are the boundary conditions used: 1) Concentration at the surface of the LB curvature = $C_s$ (saturation concentration of water vapor at 25 $^0$C and 40 % RH. 2) The concentration flux of water vapor normal to the glass slide (glass plate boundary) $\frac{dC}{dn} = 0$ where *n* is the direction normal to the glass plate. The concentration field relaxes to the relative humidity at the far-field. The far-field is considered to be at a distance of 20 mm from the liquid bridge. The mesh consists of ~37500 elements (with individual size 0.00001 mm) and 200 divisions near the LB curvature (refer to fig. 6a). The mesh is tested for grid independence, and the results do not change by varying the element size by one order. The SIMPLE algorithm with second-order discretization is used for solving the governing equation [27].

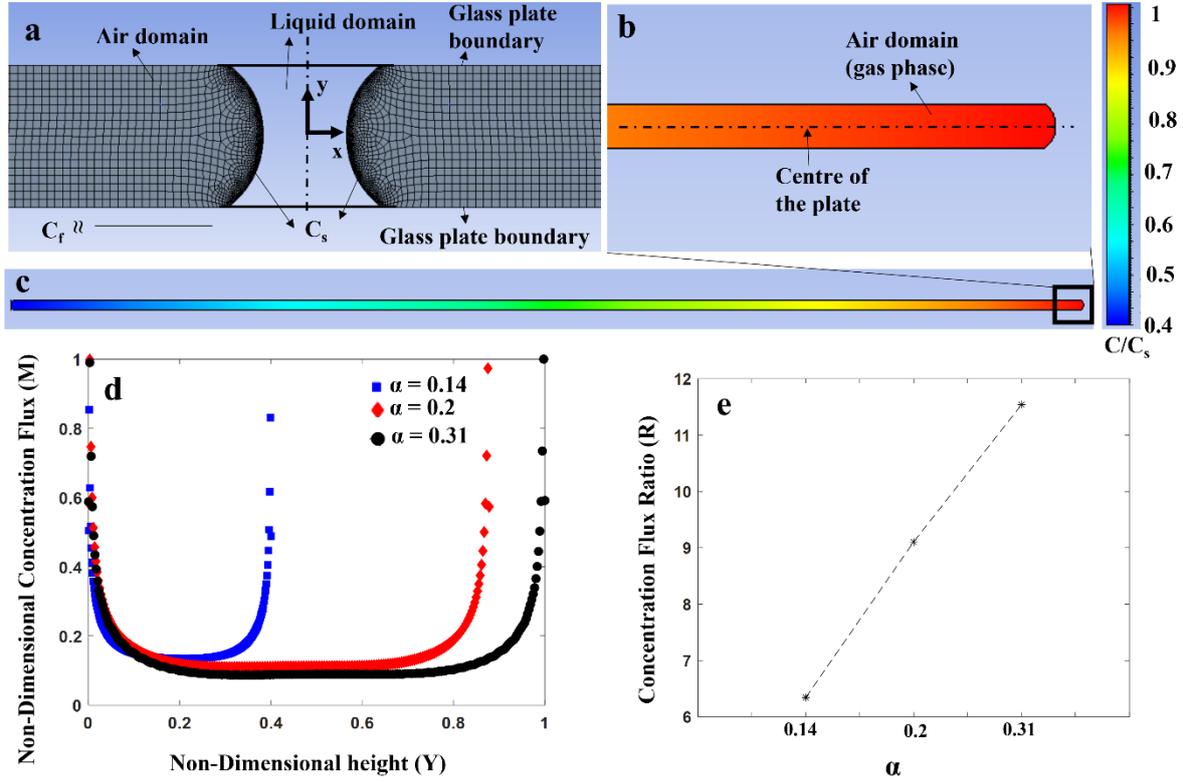

**Figure 6.** (a) Steady-state solution for the concentration field of water vapor. (a) The gas-phase domain with the meshes and boundaries. (b) the concentration countors of the water vapor near the LB. (c) the concentration countors of the water vapor in the entire domain. Plots of (d) Non-dimensional concentration flux across the surface of the LB for different values of α, (e) the ratios of concentration flux near the base to the concentration flux near the neck region for different values of α.

The concentration contours represent the solution for the governing equation in figs. 6b and 6c. The solution is similarly calculated for different values of α (= 0.14, 0.2, and 0.31). The objective is to consider at least one confinement distance from different regimes (low, intermediate, and high) and compare their performances. We non-dimensionalize the data for concise representation in a flowing manner:

1) Non-Dimensional Concentration Flux (CF) - (M)

$$M = \frac{\left(\begin{array}{c}CF \text{ on the curvature of LB} -\\ \text{Minimum } CF \text{ on the curvature of LB}\end{array}\right) for\ given\ \alpha}{\left(\begin{array}{c}Maximum\ CF \text{ on the curvature of LB} -\\ \text{Minimum } CF \text{ on the curvature of LB}\end{array}\right) for\ \alpha=0.31} \qquad (2)$$

2) Non-Dimensional height (Y)

$$Y = \frac{y\ coordinate\ on\ the\ LB\ curvature}{(\Delta h)_{\alpha=0.31}} \qquad (3)$$

3) Concentration flux ratio (R)

$$R = \frac{Water\ Vapor\ CF\ at\ the\ base\ of\ LB}{Water\ Vapor\ CF\ at\ the\ neck\ region\ of\ LB} \qquad (4)$$

It is apparent from fig. 6d that the area under the curve increases with the increase in the value of α. This corresponds to increased total evaporation with an increase in ΔH. However, interestingly the value of R (i.e., the relative concentration flux from the base of the LB to the neck region) also increases substantially (up to ~ 40% increase in R for change in the value of $\alpha$ from 0.14 to 0.31). Due to the larger gradient of concentration across the surface of LB for $\alpha = 0.31$ compared to $\alpha = 0.14$, we expect the EDF (which would be driven radially outwards) to be stronger for the case $\alpha = 0.31$. This can be experimentally verified by observing the flow inside LB for the cases $\alpha = 0.14$ and $\alpha = 0.31$.

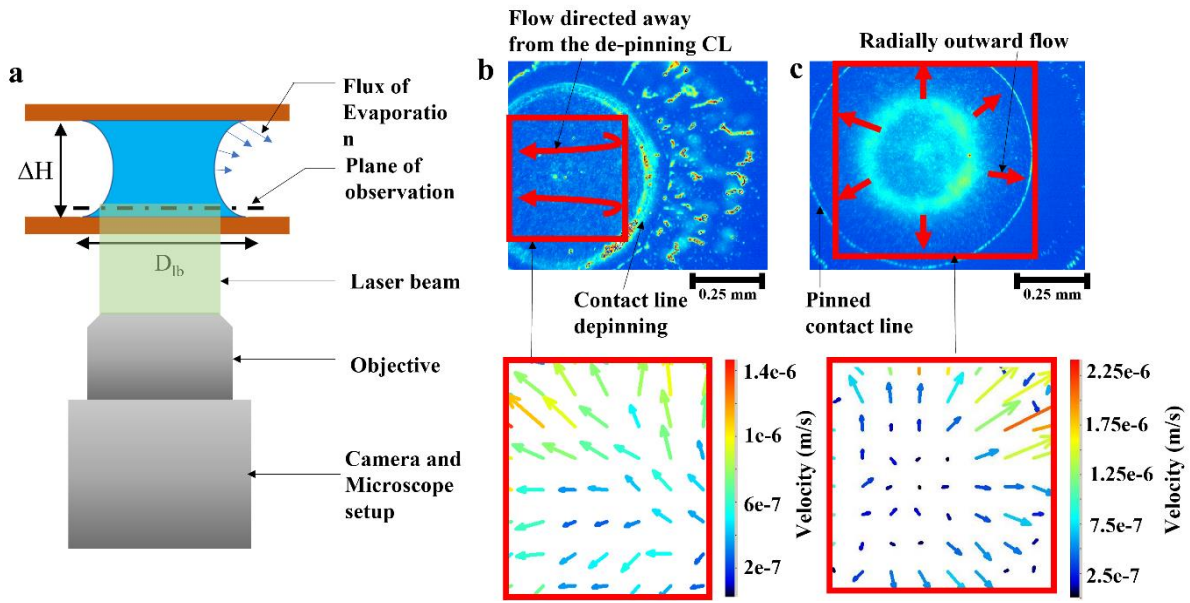

**Figure 7.** Microparticle image velocimetry (μ-PIV) for flow quantification. (a) μ-PIV set-up used in the experiments. The dashed line is the plane of observation where the images are acquired for μ-PIV. The flow visualization (arrows show the flow direction) and the corresponding averaged vectors of velocity respectively on the observation plane in the LB (b) for α=0.14, (b) α=0.31. The velocity vectors are averaged for 50 frames with starting frame of capture within 20 seconds of liquid bridge formation.

Microparticle image velocimetry (μ-PIV) is done to quantitatively assess the flow inside the liquid bridge. The liquid bridge is illuminated volumetrically by a laser beam (Nd: Yag laser NanoPIV, Litron Laser) from the bottom (refer to fig. 7a). The images are captured from the bottom in the plane of observation near the base of the LB. Imager Intense CMOS camera (LaVision$^{TM}$) fitted to a Flowmaster MITAS microscope [field of view (FOV): 1000 × 600 μm$^2$, depth of field: 28 μm] is used for capturing images. Images are acquired at the plane 50 μm above the base of the LB at 0.5 fps using the single frame-single pulse technique. It has been ensured that 3 –4 pixels of particle shift are maintained between the subsequent frames, which is deemed optimal for computation.

The gradient of evaporation (R) for case α=0.14 is significantly less, inducing low evaporation-driven flow velocity (~ $U_C$ ~$10^{-7}$ μm/s). The low radially outward flow velocity fails to compensate for the efflux of evaporation from the CL, leading to early depinning of the drop. Later the flow inside the LB is driven by the depinning force exerted by the CL (dominant force) compared to the radially outward EDF [12]. The depinning of CL for the case α=0.14 directs the flow away from the depinning contact line with a velocity scale of $U_C$ ~1 μm/s (refer to fig. 7b and video 1). However, for case α=0.31, the flow is directed radially outward towards the pinned edge with $U_C$ ~1-2 μm/s (refer to fig. 7c and video 2). As explained earlier in this section, the strength of EDF is directly dependent on the differential evaporation flux of the curvature. The higher gradient of evaporation (R) across the surface of the LB for case α=0.31 drives the flow towards the edge of the LB (similar to the capillary flow in a sessile droplet). As a result, the particles move towards the rim, permanently pinning the CL.

3E. Mechanism of particle deposition: role of thin film

From the above discussions, it is clear that evaporation characteristics are indeed different when the confinement length of the LB is altered. However, this is not sufficient to explain the differences in deposition patterns. The coffee ring patterns emerge as the droplet CL remains pinned and the particles jam into the three-phase CL due to the capillary flow. The transport of particles hinders further dewetting of CL, thereby forming a ring-like structure. However, we need to understand the rationale for particle deposition like spokes (ordered and aligned) instead of the coffee rings. To elucidate this departure from conventional coffee ring patterns, we resort to the dynamics of liquid thin film near the three-phase contact line where the particles deposit onto the surface. First of all, we anticipate that a liquid thin film of height '$h$'

(that exists till the final moment of drying during dewetting of CL) is responsible for the deposition of particles in an aligned manner, in the case of the LB droplet (fig. 8a). We also suggest that the height of the confinement distance ($\Delta H$) is proportionally related to the height of this thin film (TF). This means that cases with higher $\alpha$ will have higher $h$, and more squeezed LB will have lower $h$. Now there are three possibilities when we compare TF length scale ($h$) with the length scale of the particle ($D_p$ – diameter of the particle): (i) $h > O(D_p)$ [TF height has higher order than particle size], (ii) $h \sim O(D_p)$ [TF height and particle size are of equivalent order], and (iii) $h < O(D_p)$, [TF height is of a lower order than particle size]. We speculate that ordered spokes-like patterns are possible only when $h \sim O(D_p)$ or $h < O(D_p)$. A higher value of $h$ suggests the augmented propensity of multilayer particle agglomeration rather than monlayer ordered arrangement.

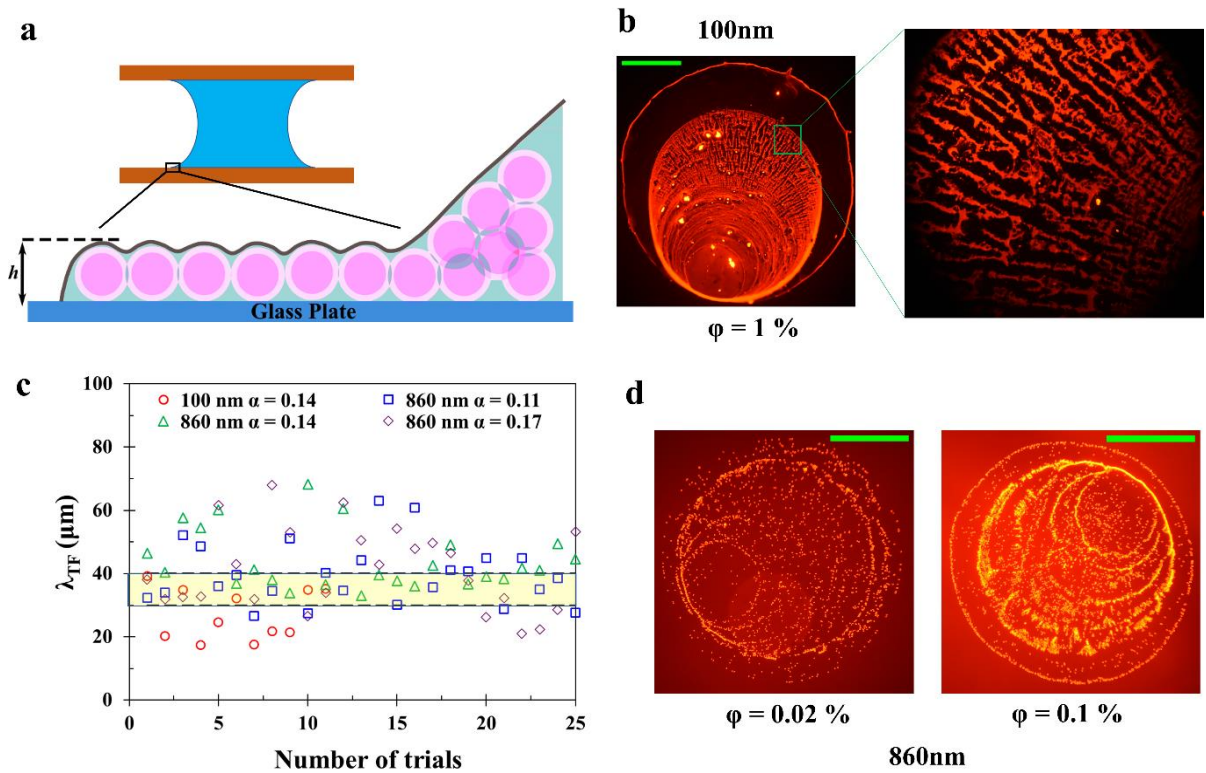

**Figure 8.** (a) Schematic figure showing deposition of particles via drying thin film near three-phase CL. (b) Deposition patterns of 100 nm colloids (concentration 1% v/v) for low confinement distance ($\alpha = 0.11$), recorded at 10× magnification (zoomed view (50×) of selected location). (c) Comparison of the theoretical prediction of characteristic length scale with the experimental observations. The yellow shaded region between two horizontal dashed lines indicates the theoretical prediction, and the different symbols correspond to experimental

observations. (d) Drying precipitates of 860 nm colloidal dispersions at two different concentrations (0.02% v/v and 0.1% v/v, both at 10× magnification) for a given confinement distance ($\alpha = 0.14$). Scale bars, indicated by a green straight line, are equivalent to 500 µm.

To check the validity of our proposition, we further investigate the patterns of colloidal LB having a smaller particle size of 100 nm (fig. 8b) with a volume fraction of 1%, maintaining a confinement distance of $\alpha = 0.11$. Ordered alignment similar to spokes was observed, although the deposition seems to be pretty dense compared to the $\alpha = 0.11$ case of 860 nm particle size (fig. 3a). This is understandable as the number density of particles is considerably higher for smaller particle sizes for the given particle concentration. This gives an idea about the length scale of this TF, and it is expected that $h \leq O(10^{-7})$ m (considering our assumptions are true, i.e, $h \leq O(D_p)$). The analysis is further supplemented by comparing the characteristic length scale obtained from theoretical relations with the experimental ones. As the thickness of the TF becomes smaller than 1 micron ($h < 1$ µm), intermolecular and capillary forces may dominate the fate of the thin film. Following this reason, mentioned by Vrij [28], we can find out the theoretical characteristic length scale from the following expression

$$\left(\lambda_{TF}\right)_{th} = h^2 \left(\frac{4\pi^3 \sigma_{sl}}{A}\right)^{1/2} \tag{2}$$

where $(\lambda_{TF})_{th}$ indicates the theoretical characteristic length, $h$ denotes the height ($10^{-7}$ m) of liquid TF, $\sigma_{sl}$ represents the surface tension (62.8 mN/m) of colloidal dispersions at the solid-liquid interface, and $A$ is referred to as Hammaker constant ($10^{-19}$ J). The experimental data (symbols) is found to be in good agreement with the theoretical predictions (shaded yellow region) as both of them are of the same order irrespective of particle sizes and confinement distances (fig. 8c).

Now the question remains whether particle concentration has any effect in the dried colloidal patterns for a given confinement distance ($\alpha = 0.14$) and given particle size (860 nm). As observed from fig. 8d, reducing the volume fraction by two orders to 0.02% results in scattered distribution of particles. However, in the case of intermediate particle concentration (0.1% v/v), disjointed spokes appear with intermittent rings. Therefore, it can be inferred that deposition patterns are dependent on the particle concentration of the colloidal media and obtaining a desired pattern can be exercised by tuning the volume fraction of the particulates. Another way to distinguish these patterns is by comparing the velocity scale of dewetting (fig. 5b). The low dewetting velocity (of the order of few nm/s, in case of $\alpha$ = 0.31, 0.34) is a signature of pinned CL, thereby leading to coffee ring deposits (refer to fig. 9a). Even when the CL slips and sticks to a new location, the existence of multiple coffee rings signifies that the timescale between two successive slips is sufficiently higher than the particle residence timescale, thereby allowing particle aggregation at the CL. However, if the velocity of dewetting is relatively higher (for $\alpha$ = 0.11, 14), or for instance, colloidal droplet goes through quasi-continuous dewetting, lesser particles have the opportunity to settle within this TF, instead of being accumulated at the CL. This may plausibly lead to an ordered alignment (refer to fig. 9b) rather than the conventional coffee ring. Further, we confirm the sequential deposition of the nano-particles for particle sizes of 860 nm and 100 nm from the SEM images (figs. SF1a and SF1b). Monolayer deposits are observed in the case of 860 nm LBs, whereas 100 nm case shows the possibility of two-three layers of deposition, confirmed from the profilometry data (fig. SF1c).

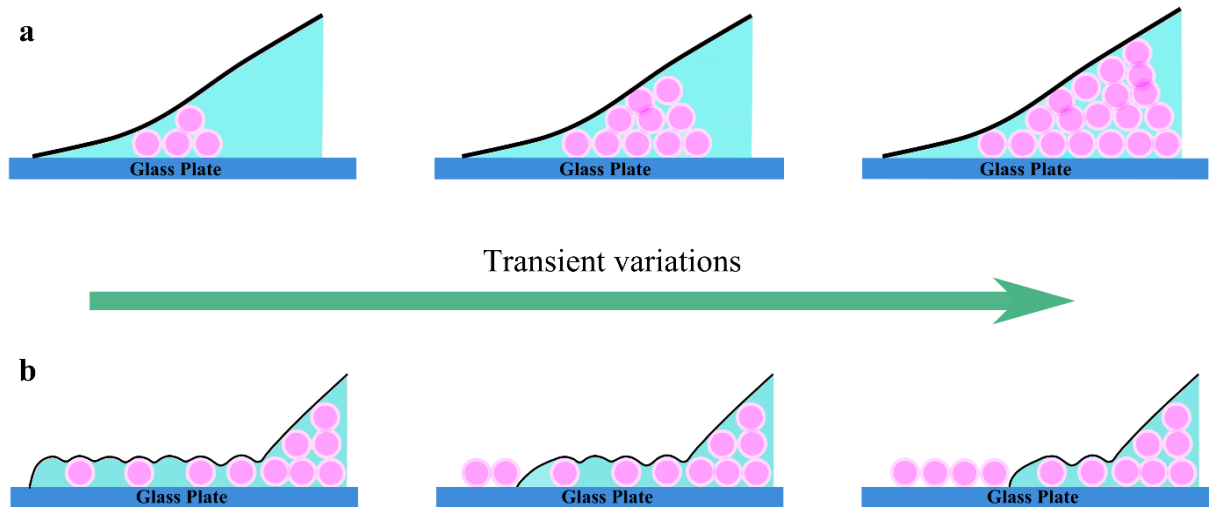

**Figure 9.** Schematic depicting the particle deposition mechanism leading to coffee ring (a) and ordered alignment similar to spokes (b).

3F. Experimental validation with reflection interference microscopy

Further, we carried out reflection interference microscopy to validate our proposed theory of liquid thin film (TF). However, before addressing the experimental observations from this TF during the dewetting process, we first present the analysis of the drying liquid film that is noticed at the bottom surface when the sessile drop initially translates into an LB. As mentioned earlier (section 3A), the sessile drop placed on the bottom glass surface transforms to LB when the top surface comes in contact with it. During this formation of LB, the exchange of the liquid layer to the top contact surface results in the reduction of the pinned surface region of the droplet in the bottom surface plate. The rapid alteration in the pinned surface region leaves a liquid layer around the LB. Therefore, the variation in interference patterns of this remanent liquid is captured using reflection interference microscopy using the high-speed camera at 250 frames per second, as shown in figs. 10a-10c (left column). When the top glass plate comes into contact with the sessile drop for the first time, high densely-spaced fringe patterns were observed from the concentrated liquid layer near the CL (fig. 10a). As the time elapses, the liquid layer's reduction with wide-spread fringe patterns and rupture of the thin-film layer nearer to the liquid bridge is evident. Figure 10c exhibits the evaporation sequence of the thin film layer that ruptures from the edge of the newly-pinned LB at the bottom surface and gradually moves towards the previously-pinned edge corresponding to the CL of the sessile drop. Due to the drying of this liquid layer, one can observe a scattered deposition of the nano-particles between the outer ring and central pattern, as evident from the patterns presented in fig. 3a.

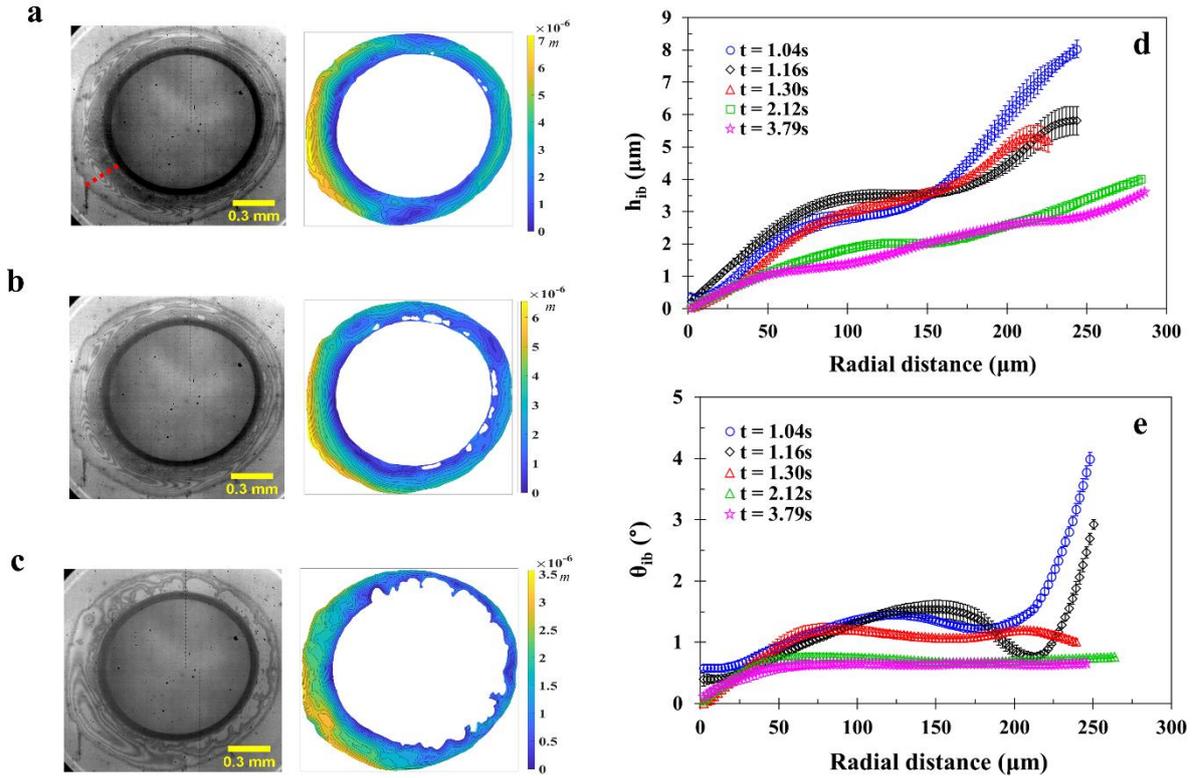

**Figure 10.** Dynamics of the liquid layer at the bottom surface when the droplet changes its configuration from sessile to liquid bridge (LB) at different time instants (a) 1 s, (b) 1.30 s, and (c) 2.21 s. Distribution of local thickness profiles (d) and contact angle (e) at the selected region (indicated by red dashed line) for various time instants.

The temporal variation in the thickness of this liquid layer and its two-dimensional distribution is determined from the variation in interference fringe patterns through employing the fast-frequency guided algorithms (refer to supplementary section). The thin-film thickness variation for time sequence is observed from the right column of the figs. 10a-10c. The two-dimensional distribution of the liquid layer exhibits the pattern of higher film thickness at the outer edges and lower film thickness adjacent to the edge of the LB, and as the liquid layer evaporates, a significant reduction in its thickness is noticed. Transient analysis reveals the average reduction of the height of this layer is ~ 8 µm to 3 µm (fig. 10d) across the selected plane, indicated by the red line in fig. 10a. The local variation of the contact angle (CA), presented in Figure 10e, demonstrates several interesting features. The contact angle measured in this analysis is different from the three-phase contact angle. As shown in fig. 10e, the CA varies spatially, whereas the three-phase CA is measured at a singular point. Here, the change in CA of the liquid layer during the evaporation is accessed through the thickness of the liquid film in the radial direction (refer eqn. 4 of supplementary). For the initial time instants, the CA is relatively higher (~ 4°) at the outer-edge periphery and follows the same trend depicted by the thickness of the film. The CA remains spatially invariant for a large region and attains a lower value (~ 1°). Further solvent depletion in the liquid layer results in almost constant CA (less than 1°), consistent with fig. 10d.

Now let us rewind the discussion of liquid TF that we have estimated to be present during the dewetting of CL contributing to the particle deposition (section 3D). Figs. 11a-11c illustrate the dewetting dynamics of the TFs of respective LBs at different confinement distances. The interference patterns were captured at 60 fps by employing a 20X magnification objective lens focusing at one of the leading edges of LB. The evolution of the receding TF for the LB of $\alpha = 0.17$ is presented in fig. 11a. With the progression of time (i.e., $t_{lb} = 57s$), the CL is observed to slip, leaving a trace of thin film between subsequent pinning events. This TF can be distinguished by observing widely-spaced fringe patterns consisting of circular loops, whereas the leading edge has closely-spaced fringes. The extent of the spread of the TF is noted at the time intervals of 155s and 157s, including the progress of additional circular loops in the fringe patterns. The development of circular loops, to a larger extent, leads to the initiation of the dewetting process in the liquid layer. Subsequently, the TF illustrated from fig. 11b reveals the identical interference patterns for $\alpha = 0.17$. The augmented evaporation rate with an increase in confinement distance of the LB is evident from the time intervals of figs. 11a and 11b, where the complete dewetting of the liquid layer occurred at $t_{lb} = 96s$ for $\alpha = 0.2$ compared to $t_{lb} = 157s$ of $\alpha = 0.17$. In the case of $\alpha = 0.25$, the receding of the leading edge is not observed from the interferometric images (fig. 11c). The transient data indicates the reduction in densely-spaced fringe patterns till the droplet remains in LB configuration. For higher confinement distances, we do not observe any dewetting of TF by the interferometric measurement, as the CL remains pinned. Several times, we had attempted to measure the TF dynamics for low confinement distances ($\alpha < 0.17$); however, we are unable to resolve the images due to infrastructural limitations of the present set-up.

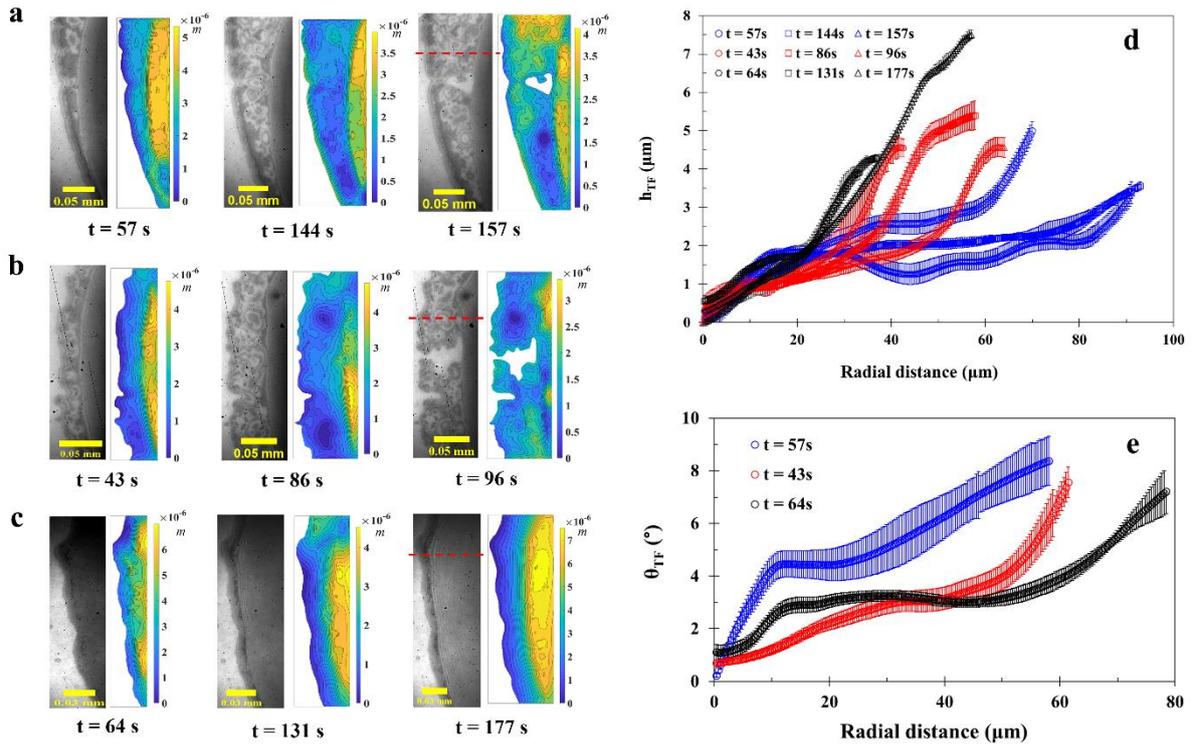

**Figure 11.** Transient variations of TFs for different confinement distances: α = 0.17 (a), α = 0.2 (b), α = 0.25 (c). Quantitative analysis of the corresponding data of figs. a, b, c is presented in terms of varying local height of the thin film (d) and contact angle (e) at the selected regions (indicated by red dashed lines) for various time instants. In the case of fig. d, the data of three different time instants (initial, intermediate, and final) of each confinement distance are denoted by different shapes of symbols (circle, square, and triangle), and the different confinement distances are indicated by symbols of blue (α = 0.17), red (α = 0.2) and black (α = 0.25) colors. Figure e shows the comparison of CAs analyzed only for initial times of three different confinement distances.

The plausible particle deposition mechanism can be explained by the thin film (TF) 's receding dynamics adjacent to the CL. In the case of α = 0.17, the development of multiple circular loops leads to multiple dewetting regions in the liquid layer due to the slower evaporation rate of LB (fig. 11a). The formation of multiple dewetting regions results in thin streak liquid films where the particles assemble sequentially in an aligned fashion during the final drying process (fig. 9). Formation of multiple slipping planes and intermittent streaks are observed for α = 0.17 case (fig. 3a). The coffee ring forms when the CL gets pinned to the surface while receding the TF gives rise to the disjointed streaks. In the case of higher confinement distances, one can expect the formation of single or multiple coffee ring patterns without observance of streaks.

The two-dimensional film-thickness profiles have been determined for corresponding the interferometric images shown in figs. 11a-11c. Figures 11d and 11e exhibit the transient variation of the height of the TF and the CA for the selected region (the respective planes indicated by the red line). For lower values of α, the elongated spread in the TF with lower film-thickness measurements was considered. The increment in confinement distances (α) of the respective LBs resulted in the enhanced average height of the corresponding TFs that reduces as the drying progresses. In particular, the TF thickness, estimated at the final instant ($t_{lb}$ = 144 s), is ~ 1 to 1.5 µm for the LB having α = 0.17. Therefore, by extrapolating the trend of fig. 11d, it can be inferred that for more squeezed LBs (lower α), the height of the corresponding TFs is expected to be at least lower than $O(10^{-6})$ m. This conforms to the observations based on our theoretical proposition, where $h$ is of submicron length scale, and we have employed the same in eqn. 2 to estimate the dominant capillary length scale. The comparison of CA (fig. 11e) shows similar behaviors as reflected in fig. 10e, where the spatial variation of CA increases towards radially inward (towards the center of LB) direction for a given confinement distance, although the magnitude remains a bit higher (~ 8°).

## 4. Conclusion:

In this article, for the very first time, we demonstrate that tailoring the confinement length of a colloidal liquid bridge (LB) can transform the post-evaporation patterns from conventional coffee ring to the patterns found on scallop shells. Altering the gap between solid surfaces modifies the curvature of the LBs, thereby giving rise to significant variations in the evaporation flux across the liquid-vapor interface. The non-uniform nature of solvent depletion contributes to the development of concentration gradient across the curvature of LB that exhibits higher differences in concentration magnitudes for increased confinement lengths. This, in turn, controls the respective flow field distributions within the LBs that hint towards the stick-slip motion of LBs at lower confinement distances. However, pinned behaviors are noticed in case of more stretched LBs. LBs corresponding to higher confinement lengths remain pinned at the solid surfaces (both top and bottom), and with time their curvature changes significantly, finally resulting in a single coffee ring. For a colloidal drop sandwiched at intermediate confinement distances, the LB exhibits dewetting of the contact line (CL) that is evident in the form of the multiple coffee rings. When squeezed to a considerable extent, the LB undergoes faster dewetting for a significant portion of its evaporation lifetime; and yields ordered deposition of particles analogous to scallop shell. During the dewetting, submicron drying liquid film near three-phase CL has been found to govern the particle deposition in an aligned fashion. The experimental observations from interferometric measurements corroborate the theoretical propositions regarding this liquid layer. Interesting to note, the gap between the solid surfaces strongly influences the thickness of the thin film. The differences in drying patterns from colloidal LBs are attributed to the compound effect of pinning/depinning of the CL coupled with evaporation-induced flow inside the LBs and the deposition of particles via drying thin film. Further, the theoretically estimated dominant capillary length scale is similar to the experimental observations, irrespective of particle size and confinement distance. The particle volume fraction is one key parameter that controls the final precipitates. This approach of generating distinct patterns by tuning the confinement lengths of colloidal LBs can be employed in various potential applications such as printing, coating, and similar areas requiring selective particle self-assembly.

**Corresponding Author**

Professor Saptarshi Basu, Department of Mechanical Engineering, Indian Institute of Science

*Corresponding author email: sbasu@iisc.ac.in

**Author Contributions**

The manuscript was written through the contributions of all authors. All authors have approved the final version of the manuscript.

Conceptualization: AC, SB; Methodology: AC, OH, SRS, SB; Investigation: AC, OH, SRS, Visualization: AC, OH, SRS, Funding acquisition: SB, Project administration: AC, SB Supervision: SB, Writing – original draft: AC, OH, SRS, Writing: editing and revision: AC, OH, SRS, SB.

SRS and OH contributed equally to the manuscript.

**Funding Sources**

SB acknowledges DRDO Chair Professorship. The authors acknowledge funding from the Ministry of Education, Government of India, and DRDO Chair Professorship.

The authors declare no competing financial interest.

**Data availability statement:** All data are available in the main text or supplementary materials. All materials and additional data are available from the corresponding author upon request.

**Acknowledgment**

The authors would like to thank Micro-Nano Characterization Facility (MNCF) at the Center for Nano-Science and Technology (CENSE), the Indian Institute of Science (IISc), for providing the facility for SEM, profilometry, and measurement of Zeta-Potential of the nano-particles.

**Reference:**

1. Sun, J., Bao, B., He, M., Zhou, H., and Song, Y., 2015. Recent advances in controlling the depositing morphologies of ink-jet droplets. *ACS applied materials & interfaces*, *7*(51), pp.28086-28099.
2. Hammond, P.T., 2004. Form and function in multilayer assembly: New applications at the nanoscale. *Advanced Materials*, *16*(15), pp.1271-1293.
3. Cao, Q. and Rogers, J.A., 2009. Ultrathin films of single-walled carbon nanotubes for electronics and sensors: a review of fundamental and applied aspects. *Advanced Materials*, *21*(1), pp.29-53.
4. Wu, L., Dong, Z., Li, F., Zhou, H. and Song, Y., 2016. Emerging progress of ink-jet technology in printing optical materials. *Advanced Optical Materials*, *4*(12), pp.1915-1932.
5. Martel, R., 2008. Sorting carbon nanotubes for electronics. *ACS nano*, *2*(11), pp.2195-2199.
6. Deegan, R.D., Bakajin, O., Dupont, T.F., Huber, G., Nagel, S.R. and Witten, T.A., 1997. Capillary flow as the cause of ring stains from dried liquid drops. *Nature*, *389*(6653), pp.827-829.
7. Yunker, P.J., Still, T., Lohr, M.A. and Yodh, A.G., 2011. Suppression of the coffee-ring effect by shape-dependent capillary interactions. *Nature*, *476*(7360), pp.308-311.
8. Xie, Q. and Harting, J., 2018. From Dot to Ring: The role of friction in the deposition pattern of a drying colloidal suspension droplet. *Langmuir*, *34*(18), pp.5303-5311.

9. Bhardwaj, R., Fang, X. and Attinger, D., 2009. Pattern formation during the evaporation of a colloidal nanoliter drop: a numerical and experimental study. *New Journal of Physics*, *11*(7), p.075020.
10. Hu, H. and Larson, R.G., 2006. Marangoni effect reverses coffee-ring depositions. *The Journal of Physical Chemistry B*, *110*(14), pp.7090-7094.
11. Parsa, M., Harmand, S., Sefiane, K., Bigerelle, M. and Deltombe, R., 2015. Effect of substrate temperature on pattern formation of nano-particles from volatile drops. *Langmuir*, *31*(11), pp.3354-3367.
12. Hegde, O., Chattopadhyay, A. and Basu, S., 2021. Universal spatio-topological control of crystallization in sessile droplets using non-intrusive vapor mediation. *Physics of Fluids*, *33*(1), p.012101.
13. Kabi, P., Pal, R. and Basu, S., 2020. Moses effect: splitting a sessile droplet using a vapor-mediated marangoni effect leading to designer surface patterns. *Langmuir*, *36*(5), pp.1279-1287.
14. Montanero, J.M. and Ponce-Torres, A., 2020. Review on the dynamics of isothermal liquid bridges. *Applied Mechanics Reviews*, *72*(1).
15. Kumar, S., 2015. Liquid transfer in printing processes: liquid bridges with moving contact lines. *Annual Review of Fluid Mechanics*, *47*, pp.67-94.
16. Lin, Z. and Granick, S., 2005. Patterns formed by droplet evaporation from a restricted geometry. *Journal of the American Chemical Society*, *127*(9), pp.2816-2817.
17. Xu, J., Xia, J., Hong, S.W., Lin, Z., Qiu, F. and Yang, Y., 2006. Self-assembly of gradient concentric rings via solvent evaporation from a capillary bridge. *Physical review letters*, *96*(6), p.066104.
18. Mondal, R. and Basavaraj, M.G., 2019. Influence of the drying configuration on the patterning of ellipsoids–concentric rings and concentric cracks. *Physical Chemistry Chemical Physics*, *21*(36), pp.20045-20054.
19. Leng, J., 2010. Drying of a colloidal suspension in confined geometry. *Physical Review E*, *82*(2), p.021405.
20. Boulogne, F., Giorgiutti-Dauphiné, F. and Pauchard, L., 2013. The buckling and invagination process during consolidation of colloidal droplets. *Soft Matter*, *9*(3), pp.750-757.
21. Mahanta, T.R. and Khandekar, S., 2018. Evaporation Characteristics of a Confined Nanofluid Bridge between Two Heated Parallel Plates. *Journal of Flow Visualization and Image Processing*, *25*(3-4).

22. Mondal, R. and Basavaraj, M.G., 2020. Patterning of colloids into spirals via confined drying. *Soft Matter*, *16*(15), pp.3753-3761.

23. Upadhyay, G. and Bhardwaj, R., 2021. Colloidal Deposits via Capillary Bridge Evaporation and Particle Sorting Thereof. *Langmuir*, *37*(41), pp.12071-12088.

24. Siddon, C.E., Smith, Q.T., McNeel, K., Oxman, D. and Goldman, K.J., 2017. *Protocol for Estimating Age of Weathervane Scallops Patinopecten caurinus in Alaska*. Alaska Department of Fish and Game, Division of Sport Fish, Research and Technical Services.

25. Bhardwaj, R., Fang, X., Somasundaran, P. and Attinger, D., 2010. Self-assembly of colloidal particles from evaporating droplets: role of DLVO interactions and proposition of a phase diagram. *Langmuir*, *26*(11), pp.7833-7842.

26. Larson, R.G., 2014. Transport and deposition patterns in drying sessile droplets. AIChE Journal, 60(5), pp.1538-1571.

27. Khawaja, H. and Moatamedi, M., 2018. Semi-Implicit Method for Pressure-Linked Equations (SIMPLE)–solution in MATLAB®.

28. Vrij, A., 1966. Possible mechanism for the spontaneous rupture of thin, free liquid films. Discussions of the Faraday Society, 42, pp.23-33.